\documentclass[preprint,3p,twocolumn]{elsarticle}
\usepackage{graphicx}
\usepackage{mathrsfs}
\usepackage{algorithm,algorithmicx, algpseudocode}
\usepackage{amsmath}
\usepackage{caption}
\usepackage{multicol}
\usepackage[export]{adjustbox}
\usepackage{lineno,hyperref}
\usepackage{fancyvrb}
\usepackage{bchart}

\modulolinenumbers[5]

\def\CircleArrow{\hbox{$\circ$}\kern-1.5pt\hbox{$\rightarrow$}}

\algnewcommand\algorithmicnot{\textbf{not}}
\algdef{SE}[IF]{IfNot}{EndIf}[1]{\algorithmicif\ \algorithmicnot\ #1\ \algorithmicthen}{\algorithmicend\ \algorithmicif}%

\journal{Journal of \LaTeX\ Templates}

\let\today\relax
\makeatletter
\def\ps@pprintTitle{%
    \let\@oddhead\@empty
    \let\@evenhead\@empty
    \def\@oddfoot{\footnotesize\itshape
         {~} \hfill\today}%
    \let\@evenfoot\@oddfoot
    }
\makeatother

\begin{document}

\begin{frontmatter}

\title{Novel data structures for label based queries specifically efficient for
billion+ property graph networks using Kinetica-Graph$^{\vcenter{\hbox{\includegraphics[height=0.35cm]{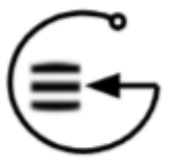}}}\dagger}$}

\author{B. Kaan Karamete\corref{cc}} 
\author{Eli Glaser\corref{}}

\cortext[cc]{\it Corresponding author: Bilge Kaan Karamete, \\ 
kkaramete@kinetica.com, karametebkaan@gmail.com\\ 
$\dagger$Kinetica-Graph: https://arxiv.org/abs/2201.02136 }
\address{Kinetica DB Inc. \break 901 North Glebe Road, Arlington, Virginia 22203}

\begin{abstract}
This paper discusses a novel data structure that efficiently implements label based graph queries particularly for very large graphs. The major issues in large graph databases is the memory foot-print of label based property associations to graph entities and subsequent query speeds. To this end, unlike the available graph databases, that use key-value pairs using map like associative containers, we have devised a novel data structure that is superior in its memory foot-print as well as its fast search characteristics without any compromise on the number of labels that can be associated to graph nodes and edges. We will demonstrate the power of this novel unconventional data structure over billion plus graphs within the context.

\noindent 

\end{abstract}


\end{frontmatter}

\section{Introduction}

It has been a common software engineering task to find the most feasible and efficient data structures/containers for the algorithms to succeed with an acceptable outcome often measured by its accuracy, storage characteristics and its speed of execution. Modern data structures such as the widely used STL~\cite{stl} containers are rectified over the past decade to do the heavy lifting and take the burden off the algorithms as much as possible. Re-usability of the data structures was the main goal so that the  algorithms could be designed for any data types. This flexibility of using the same containers provided generic solutions via the use of templates as mere space-holders for any data/object type~\cite{cppdata}. In 80s, and 90s, when the computing power was not at its apex, but certainly in its  ascendancy, often referred to as the Fortran era (no pun intended), there was more focus on the data structures to fit into smaller range of memory architectures that existed in the hardware technology at the time. Templated objects and structures were not at the fore-front for a very valid reason of being able to solve as big as possible within the existing limited comput-power. However, as the computational prowess improved over the following decades, the appetite for better and more efficient data structures did not diminish, and particularly for certain use cases where even with the advent of new hardware, the need for light and efficient data structures never ceased to exist. 

In today's fast paced software practices, habitual selection of readily available standard STL data containers are preffered with not much concern on the scalability that often lead to unacceptable results when the data size is immense and cross relations play a big factor in speed and storage. The deviation from efficiency is more pronounced particularly for billion+ large graphs in which relations between the graph entities and the unlimited string labels result in one-to-many and many-to-many associations. Many graph vendors implement the label associations to graph nodes and edges using this habitual standard key-value look-up containers~\cite{graphdbs, neo4j, tiger, networkx}. Instead of addressing the efficiency of these data-structures, these vendors suggested the distributed graph representations when it is simply not possible to store and solve large (billion+) graphs. However, suggesting a wrong recipe for a much simpler issue creates bigger bottlenecks; the clustering of the partitions has a huge impact on the speed of the distributed solves and queries since the scalar values have to be transferred across adjacent partitions many times for convergence. If the solve or query graph traversals need to jump back and forth across partitions frequently due to inter-mingling of the partitions with poor clustering characteristics, it would take many more iterations to converge. Hence the seemingly much simpler problem of devising a light weight and efficient data containers the issue is transformed into a more complex problem of how to make partitioning more efficient with the least amount of inter-partition boundary nodes. 

\begin{figure}
\centering
    \framebox{\includegraphics[width=\linewidth, keepaspectratio]{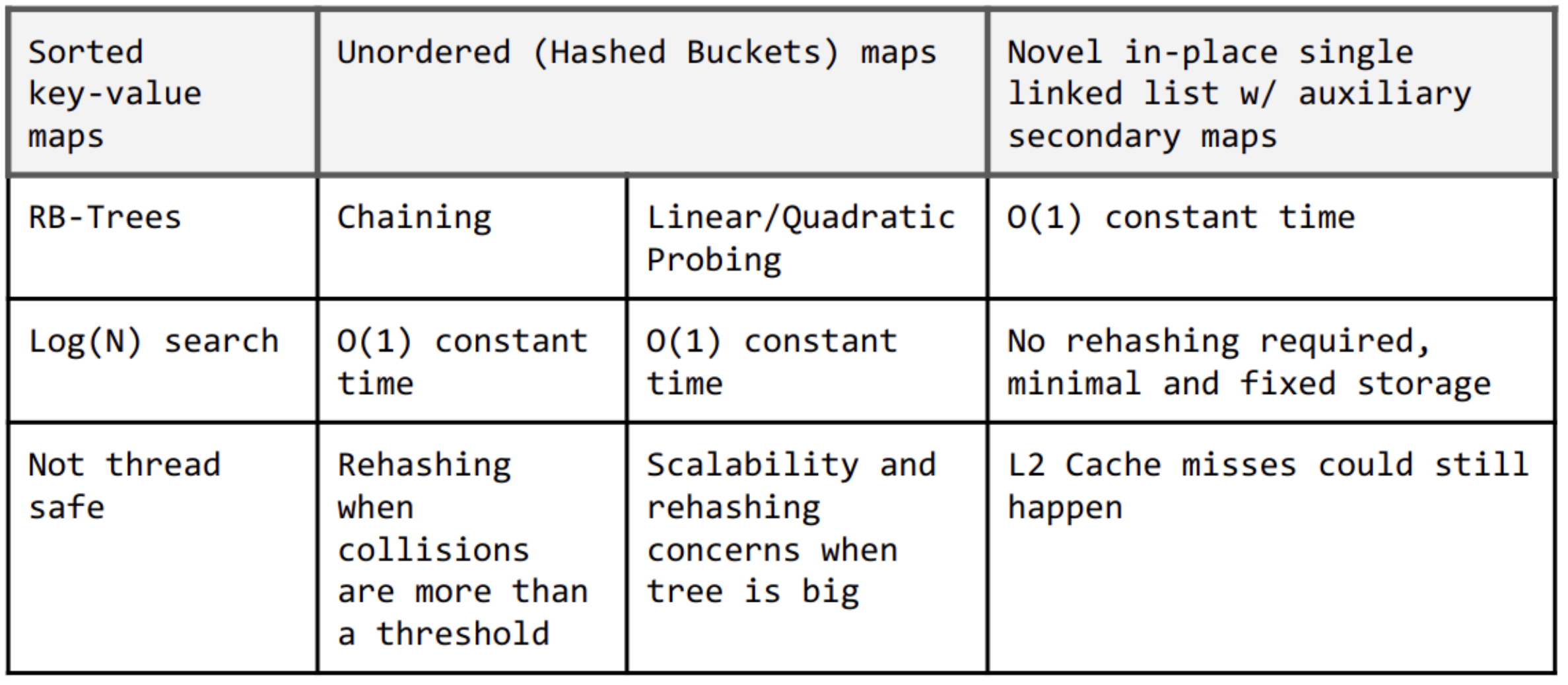}}
    \caption{The comparison chart for key-value containers - far right column is our native containers, that will be explained in the context.}
    \label{Figure:maps}
\end{figure}

Before we introduce and suggest this light weight and super efficient entity-label data structures, it is prudent to give some technical details and summary over the conventional sorted or hashed look-up containers. Singly or multiply associative key-value pair look-up containers often implement red-black trees (RBTrees) for sorted balanced trees~\cite{rbt,rbt2}, a binary search structure with user-specified sort functions or the hash tables where the item is hashed with the user specified hashing function. For the latter, the item is appended to a list of items whose hash keys are the same within the same bucket, known as hashing by chaining~\cite{hash,hash2}, or using a mechanism to spread the data, called probing~\cite{probing,probing2} by pinpointing to the empty bucket locations for the same hash-key entries (hash-collisions). Hashing by chaining technique is easier to implement and control, however, for dynamic scenarios, reallocation (rehashing) is required that can quickly result in de-fragmentation and large memory foot prints, on the other hand probing techniques could suffer on both the speed and the large memory issues for very large data particularly for high number of collisions as well. Sorted maps (RBTrees) has the logarithmic search time complexity, however, the performance quickly degrades when the tree gets larger as summarized in Figure~\ref{Figure:maps}. These are the known issues for hashed and sorted maps but for up-to a few to ten million item key-value look-ups, these data structures are usually very hard to beat on many technical merits that they offer reliably standard implementations over a variety of OS architectures~\cite{benchmark}. 

In this paper, we'll explain a very light weight and efficient novel data structures in modeling the graph labels to nodes and edges that enabled us creating $4+$ billion edge graph with $64-char$ labels per node/edge within less than a 1TByte (Terra Byte) machine as well as distributed over 6-machines.  The graph ontology along with the direct and inverse relations between entities and labels will be discussed on a simple example in Section~\ref{Section:SampleInput}. The new data structures will be covered in Section~\ref{Section:NewDataStructure}. Finally, in Section~\ref{Section:Parallelism}, a billion+ graph case will be demonstrated both on single node and distributed mode using this super efficient labeling framework.

\begin{figure}
\centering
    \framebox{\includegraphics[width=0.9\linewidth, keepaspectratio]{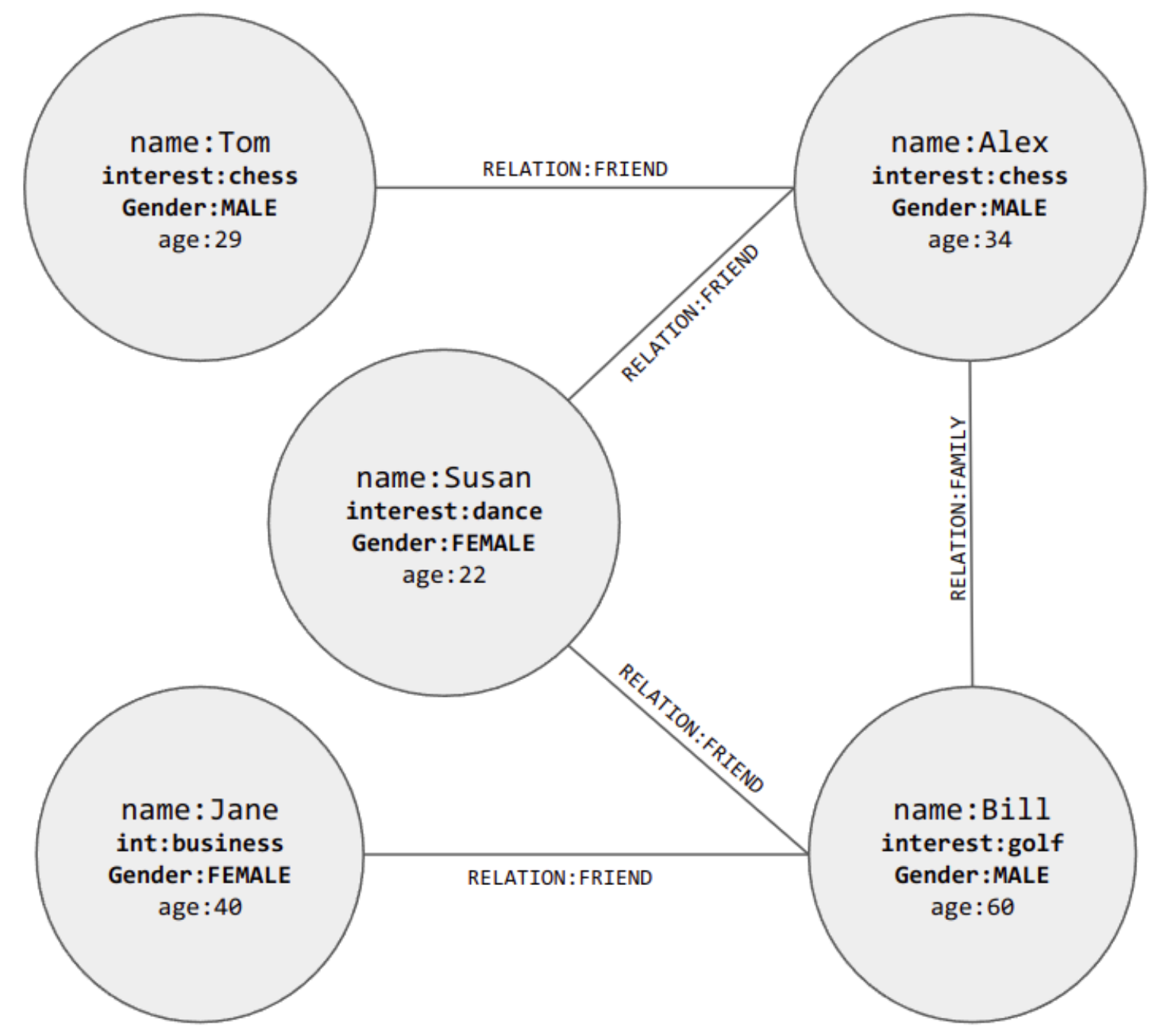}}
    \caption{This is a simple graph that is often found on wikipedia that has five (nodes) from the names, and with Gender and interest labels connected to each other via edges that have relation label. Note that the 'Age' column is not stored by the graph but accessible for graph restrictions etc, via the hybrid OLAP expression support.}
    \label{Figure:wikigraph}
\end{figure}

\section{Sample Input}
\label{Section:SampleInput}

\begin{figure}
\centering
    \framebox{\includegraphics[width=0.8\linewidth, keepaspectratio]{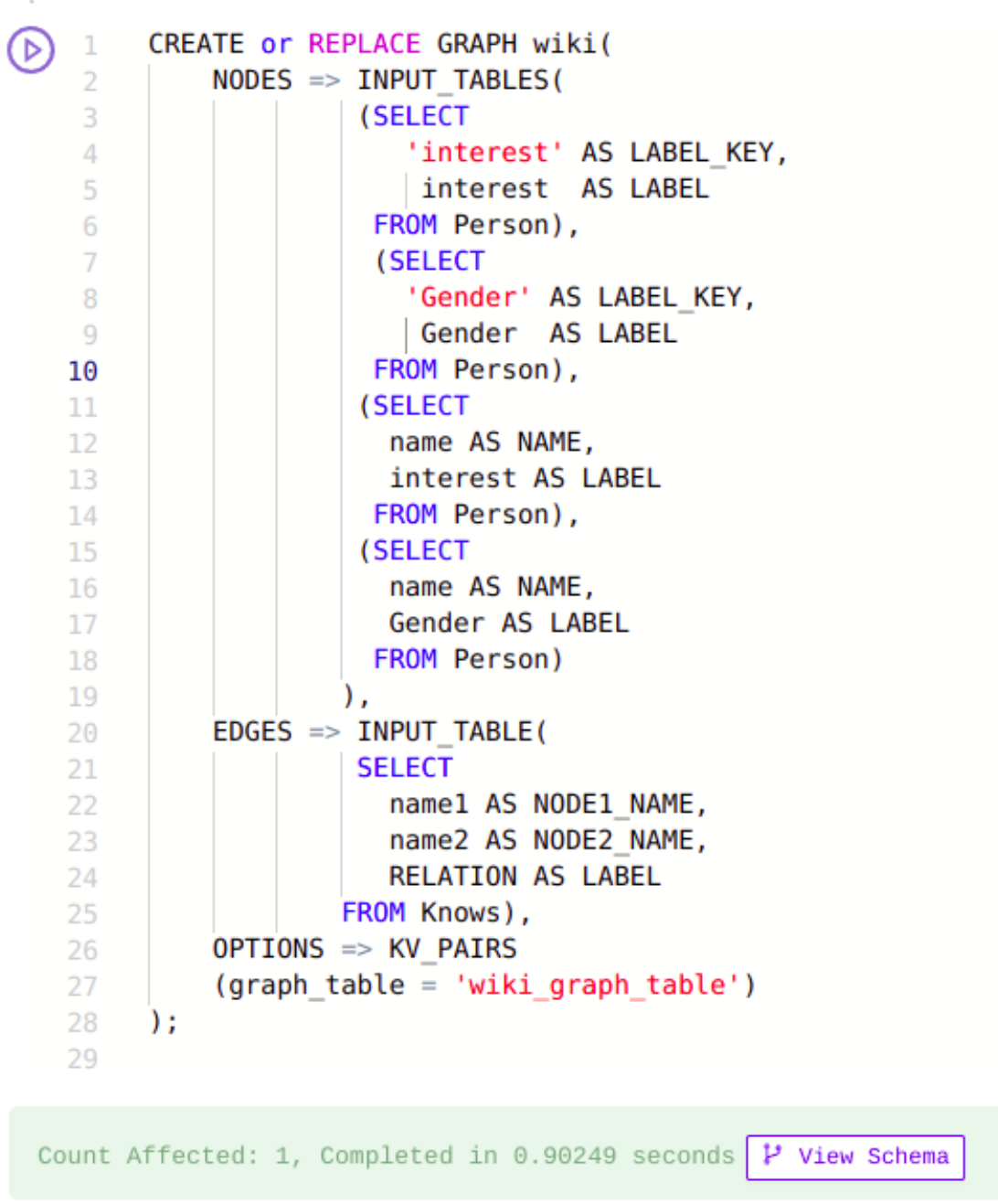}}
    \caption{Our graph SQL syntax for creating the graphs from table-column annotation combination tuples for labels, nodes, and edges. For example, the edges are specified to be constructed from the $name1$ and $name2$ columns' record values of $Knows$ table. Our adhoc graph grammar, depicted  is intitutive and extendable }
    \label{Figure:wikisql}
\end{figure}

\begin{figure}
\centering
    \framebox{\includegraphics[width=0.8\linewidth, keepaspectratio]{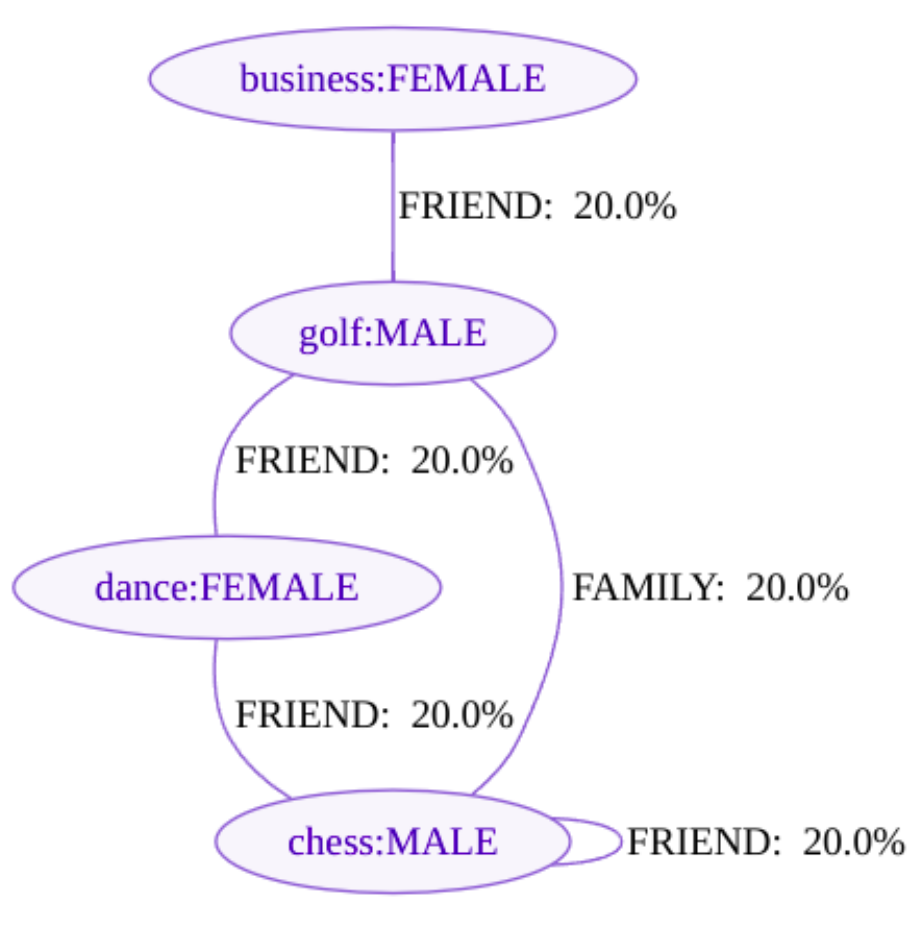}}
    \caption{This is the graph ontology generated from its node and edge labels to depict how labels are connected to each other. Our create/graph endpoint (API) generates this information as dot graph format in its response. For example, $20\%$ of the graph edges are in between \textit{chess:MALE} node labels shown as a self-looping edge above.}
    \label{Figure:wikischema}
\end{figure}

A very simple example from Wikipedia is chosen to depict the main characteristics of this new data structure 
as shown in Figure~\ref{Figure:wikigraph}. In Kinetica-Graph~\cite{kineticagraph},  we have devised an intuitive set of SQL constructs for graph endpoints. In Figure~\ref{Figure:wikisql}, corresponding SQL statement is shown for creating a directed graph where \textit{Gender} and \textit{Interest} are the labels to nodes and \textit{RELATION} column is used to depict the edge labels. We also created another sub-graph from labels to show how the labeled entities are connected to each other automatically, as the graph's ontology in the response field of the endpoint. This information is encoded as the digraph format ascii file provided by the graphviz library~\cite{graphviz} so that we could depict the result as the graph schema in our graph-UI as shown in Figure~\ref{Figure:wikischema}.

\section{New Data Structure}
\label{Section:NewDataStructure}

\begin{figure*}
\centering
    \framebox{\includegraphics[width=0.65\linewidth, keepaspectratio]{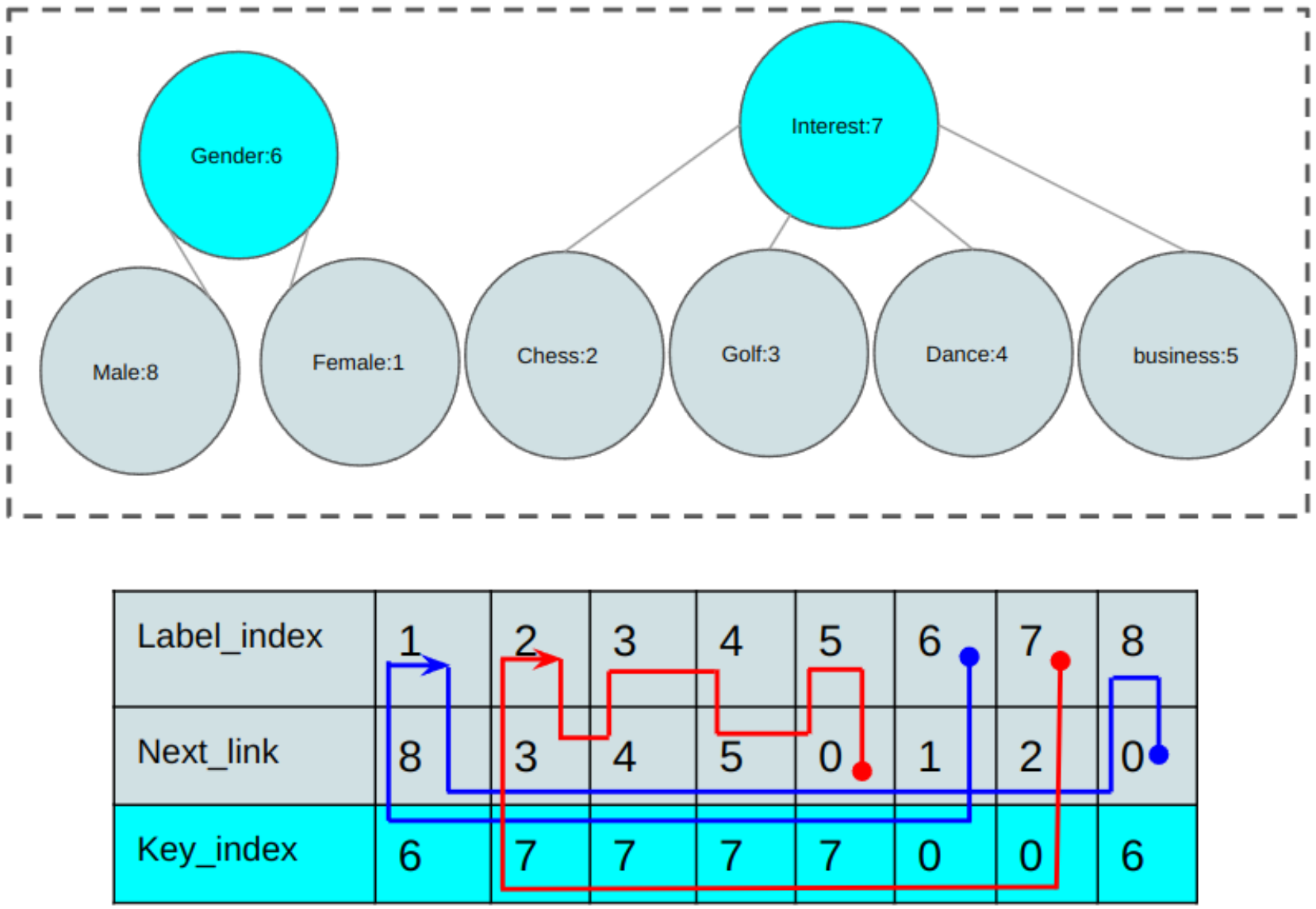}} 
    \caption{Label key $Gender=6$ is associated with the grouping for the label values of $Male=8,Female=1$ and the $Interest=7$ key has the respective label values of $Chess=2, Golf=3, Dance=4, Business=5$. The data is stored in a novel 2-item vector (bottom) where label index is the vector index and it points to the next node that the node is associated with the same key. Note that keys and label values share the same index space, i.e., keys and labels should be unique strings across both sets as the design constraint.}
    \label{Figure:labelset}
\end{figure*}

Proposed labeling framework has a few moving pieces:
\begin{itemize}
\item An \textbf{unordered map} whose key is a set of sorted label indices and value to a unique label set tuple index. For example: say, there exists a node that is associated to both \textit{MALE} and \textit{chess} labels, then the corresponding dict encoded integers to these strings, say, $8$ and $2$, respectively, are fused as one set being $\{8,2\}$ is mapped to a unique tuple index of, say, $1$ (See Figure~\ref{Figure:labelset}), referred as $Labels2index$ in Algorithm~\ref{Algorithm:addvertex}.
\item The inverse of above, i.e., a \textbf{vector} whose index corresponds to the label set tuple index and value to the set of its label indices i.e., $1$ to $\{8,2\}$, referred as $Index2labels$ in Algorithm~\ref{Algorithm:addvertex}.
\item A \textbf{vector} for each label showing which label set tuple index it belongs to. \textit{MALE}, i.e., $8$ to unique tuple index $\{1\}$ and similarly, \textit{chess} as $2$ to the same unique tuple index $\{1\}$. As either label is fused with another label combo these entries are updated. For example, $\{MALE,golf\}$ appears with the node \textit{Jane}, and hence, a new tuple index of $4$ has to be added to the \textit{MALE's} corresponding vector of tuple indices, as $\{1,4\}$, referred as $Label2indices$ in Algorithm~\ref{Algorithm:addvertex}.
\item \textbf{DLS structure}~\cite{dls}, i.e., an in-place single link data structure between the tuple indices and the graph entity (node/edge) index  (see Figure~\ref{Figure:labelnode}): 
\begin{itemize}
 \item For each tuple index, there is a cached entity index - the size of this \textbf{vector} is as big as the number of label tuple combinations.
 \item For each entity index, there are two values; the tuple index the entity belongs to and the next graph entity index that also belongs to the same tuple  index. The size of this \textbf{vector} is two times the number of graph nodes/edges.
\end{itemize}
This structure is referred as $SingleDLS$ in Algorithm~\ref{Algorithm:addvertex}.
\item There is an automatic recycling of indexes: A \textbf{FIFO queue} created for dangling tuple indexes - where there is no graph entities associated with that index so that the same tuple index would be re-used for new label combinations. This is important as it limits the size of the vectors used above. An integer id is also stored and incremented each time to generate a new index id when the recycling queue is empty, referred as $Recyle$ and $Maxid$, respectively, in Algorithm~\ref{Algorithm:addvertex}.
\end{itemize}

The above data structures have their own serializations for byte dump into the cold or disk persists. Adding a label to a graph entity algorithm is provided in Algorithm~\ref{Algorithm:addvertex} and Algorithm~\ref{Algorithm:addlabel}. Basically, the entity's associated set of labels, i.e., a tuple index corresponding to this set is found, and if there was no label associated already, a tuple index is created for the single label, then the tuple index is inserted into the single linked list of the entity so that there is a in-place link list among the entities that have been associated with the same set of labels. New tuple index generator and recycling structures are also updated within the procedure illustrated in Algorithm~\ref{Algorithm:addlabel}. For more detailed understanding of the algorithm and the data structures for the new labeling framework please visit the publicly available github project composed of two header files only. $SingleDLS$ data-structure will be explained next in Section~\ref{subsection:singlelist}.

\subsection{\textbf{In-place single linked list}}
\label{subsection:singlelist}

Note that the key design concept of the in-place single linked lists, which is also used as the graph edge topology data-structure in Kinetica-Graph, is that there are two quantities, the entity and an associated index in this case and there can be many entities associated with the same index, but once an entity is associated with an index it can not be multiply associated with another index. Otherwise, in-place book-keeping of the linked ring of entities that belong to the same index can not be conceptualized. So, the key concept in devising this new labeling framework has been to satisfy this strong unique associativity condition, which required us to come up with the new concept of unique indexes corresponding to sets of labels to create what is so called a tuple index. If there is only one label to be attached to an entity, then the tuple has one item in the list, and if another label to be attached, then a new unique tuple index is to be created from the set of two labels. 

\begin{figure}
\centering
    \framebox{\includegraphics[width=\linewidth, keepaspectratio]{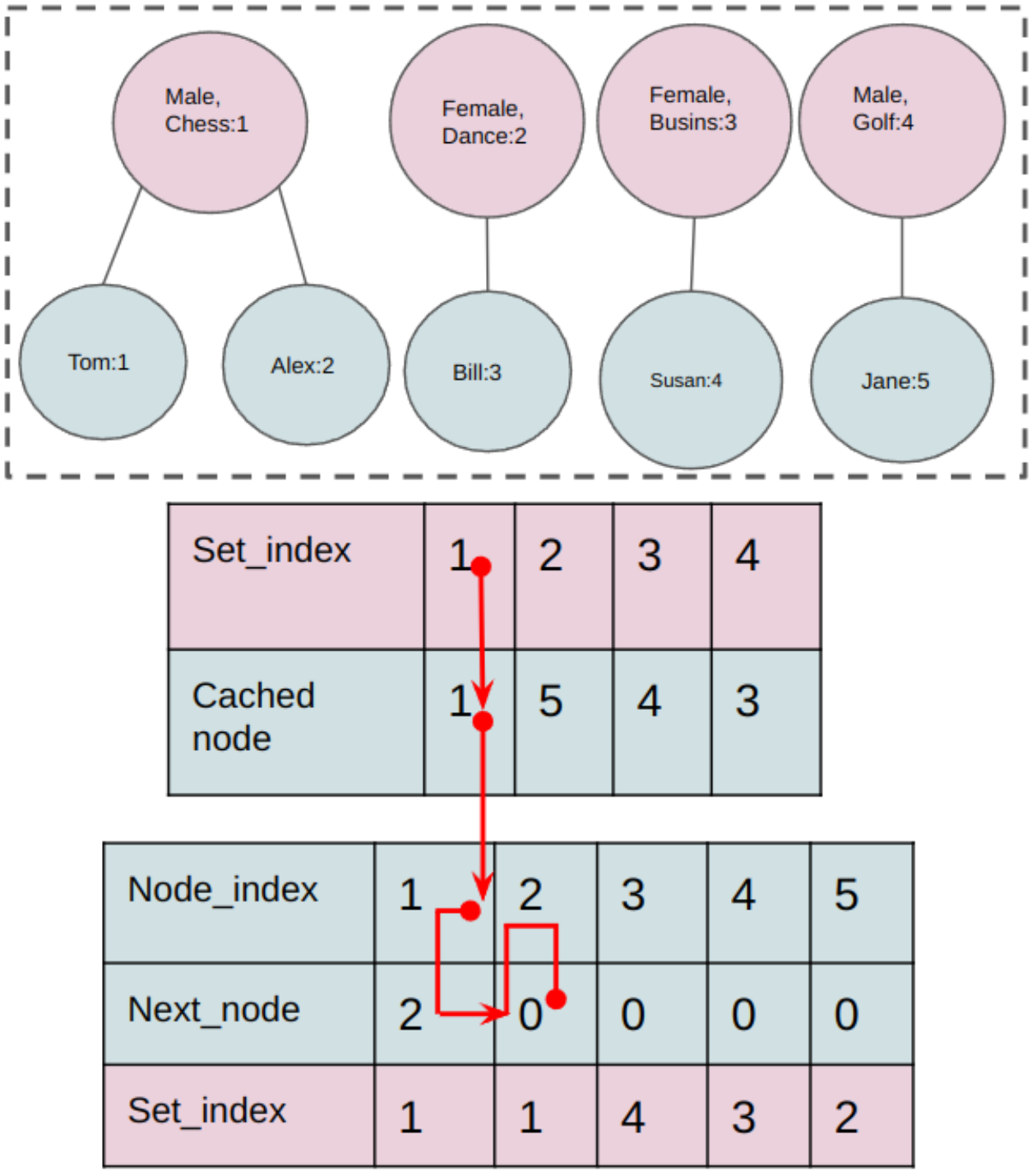}} 
    \caption{Label set index $1$ is composed of a label indices set of ${2,8}$ for $Chess=2$ and $Male=8$, respectively. This index is associated with the Cached graph node of $1$ which is $Tom$ and via the in-place single list from Node index $1$ to $2$ as $Alex$.}
    \label{Figure:labelnode}
\end{figure}

In order to illustrate this, a key-value label grouping data-structure is shown in Figure~\ref{Figure:labelset}. For instance, the \textit{Gender} is a label classifier (key-label) for label values of \textit{Male} and \textit{Female}. Likewise, \textit{Interest} is another key-label that four sub-interests are created under this category as value-labels. The in-place single link list for the label key and value pairs are embedded  in one vector as illustrated; e.g., all the interest labels can be unraveled by following the $Next\_link$ from the index of the $Interest$ key-label of $7$, which goes to index $2$, and the $2^{nd}$'s $Next\_link$ of $3$ is traversed next until the ring stops at the $5^{th}$ ($business$) value-label (see the red lines), forming a set of value labels $\{2,3,4,5\}$ for the key-label index $2$ (\textit{Interest}) as shown in Figure~\ref{Figure:labelset}.  The same concept is used in the $SingleDLS$ where entity vs label tuple index associations are stored as shown in Figure~\ref{Figure:labelnode}. The single unique association of  a tuple index (corresponds to a set of label indices) to a number of entities made this key idea possible to use for the graph labels. This light weight data-structure is very efficient since it has a fixed storage and random access search characteristics, which is a game changer when there are unlimited number of associations are sought for very large size graphs.

\begin{algorithm*}
\caption{Adds the string label and associates it with the vertex entity}
\begin{algorithmic}[1]
\Procedure{AddVertexLabel}{$Graph, Vertex, Label$}
\State $label\_index \gets Graph.GetDictEncodedIndex(Label)$
\State $pair\_index \gets SingleDLS.get\_pair\_index(Vertex)$
\State $old\_index \gets pair\_index$
\IfNot{$pair\_index$}
    \State $pair\_index \gets addLabel(vector(1,label\_index))$
\Else
    \State $existing\_tuples \gets Index2labels[pair\_index]$
    \IfNot{$binary\_search(existing\_tuples, label\_index)$}
      \State $newpair \gets existing\_tuples$
      \State $newpair.insert(upper\_bound(newpair,label\_index),
      label\_index)$
      \State $pair\_index \gets addLabel(newpair)$
    \EndIf
\EndIf
\State $SingleDLS.insert(Vertex, pair\_index)$
\If{$old\_index~and~old\_index~\neq~pair\_index)$}
   \State $recycle(old\_index)$ \Comment{See github source; recycle() method.}
\EndIf
  \State \Return $pair\_index$
\EndProcedure
\end{algorithmic}
\label{Algorithm:addvertex}
\end{algorithm*}

\begin{figure}
\centering
    \framebox{\includegraphics[width=0.9\linewidth, keepaspectratio]{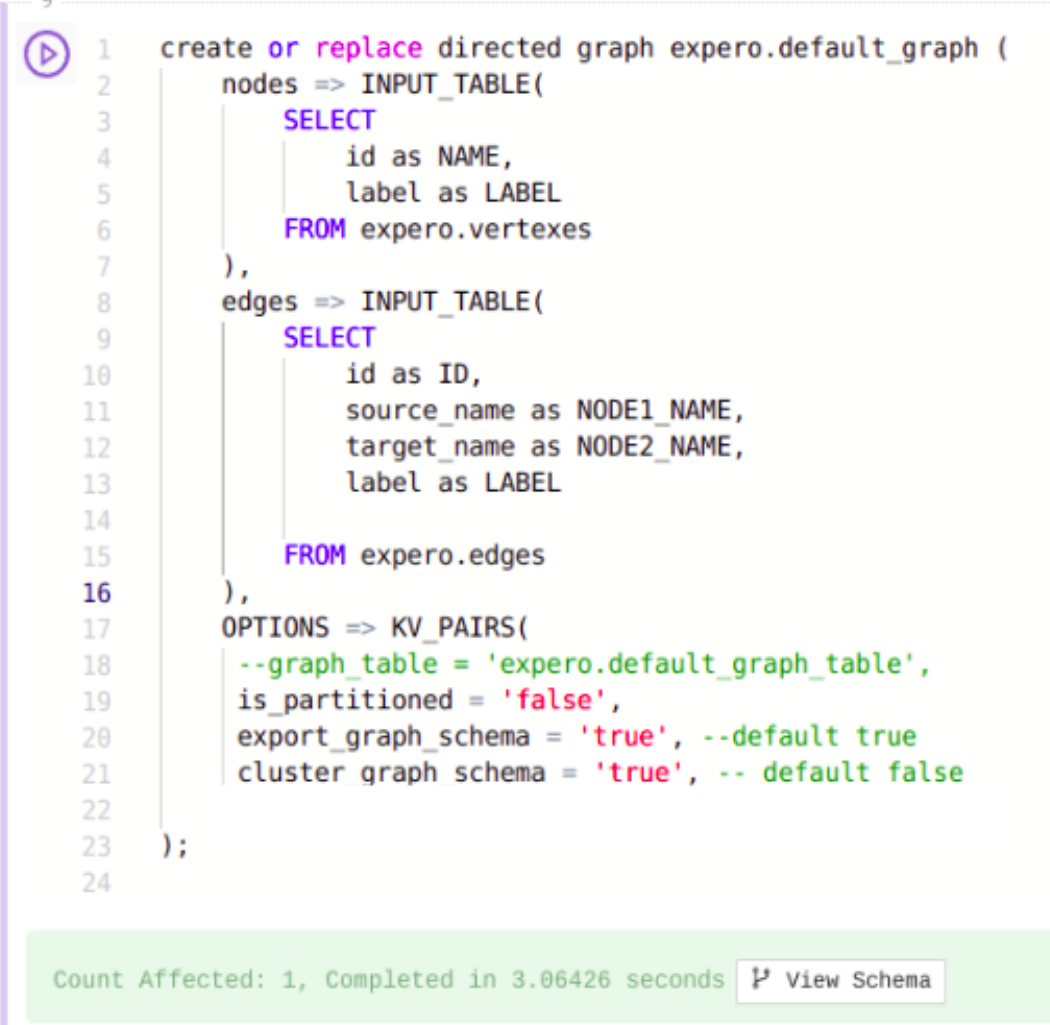}} 
    \caption{Expero graph generated using the SQL create/graph statement from the nodes and edges tables. The nodes are char64 strings and one or more labels is attached to each node and edge. There are 16 node labels, and 34 edge labels in this dataset. The graph size is 4.3 billion edges with 2.8 billion nodes. The results of the above SQL call is shown in Figure~\ref{Figure:distributed} for both distributed and single cases. The graph schema (graph ontology) as the graph of labels is shown in Figure~\ref{Figure:ontology}.}
    \label{Figure:creategraph}
\end{figure}

\begin{algorithm*}
\caption{Adds a set of labels to the framework and returns the pair index corresponding to this set}
\begin{algorithmic}[1]
\Procedure{AddLabel}{$vector~newpair$}
\State $next\_index \gets Maxid+1$
\IfNot{$Recycle.empty()$}
  \State $next\_index \gets Recycle.back()$
\EndIf  
\State $Iterator \gets Labels2Index.Insert(\{newpair, next\_index\})$
\State $pair\_index \gets iterator.first.second$ 
\If{$iterator.second$} \Comment{If the newpair index is indeed a new tuple set}
    \IfNot{$Recycle.empty()$}
          \State $Recycle \gets Recycle.Size()-1$
    \Else
          \State $Maxid \gets Maxid + 1$
    \EndIf        
    \For{$label\_index$ in $newpair$}
       \If{$label\_index \ge Label2indexes.Size()$}
          \State $Label2indexes \gets Label2indexes.Size() + 1$        
       \EndIf
       \State $Label2indexes[label\_index].Insert(upper\_bound(Label2indexes,pair\_index),pair\_index)$
    \EndFor
    \If{$pair\_index \ge Index2labels.Size()$}
       \State $Index2labels \gets Index2labels.Size() + 1$
    \EndIf
    \State $Index2labels[pair\_index] \gets newpair$         
\EndIf   
\State \Return $pair\_index$
\EndProcedure
\end{algorithmic}
\label{Algorithm:addlabel}
\end{algorithm*}

\section{Single and Distributed modes}
\label{Section:Parallelism}

The labeling framework explained in the previous Section~\ref{Section:NewDataStructure} is scalable, fixed and small, enabling queries in constant time. Hence the memory bottle-neck for large graphs is circumvented to a degree that does not require to distribute the graph at least for upto ten-twenty billion node/edges within 1-2 TByte RAM limits. We have also exercised a built-in transparent compression utility called ZRAM~\cite{zram}, that compresses the large in-place single list vectors implicitly and unzips when the data is needed, all behind the scenes in real time, with a very acceptable performance penalty increase of $20\%$ based on our findings. It is possible to get $1\colon4$ compressions with ZRAM that enables us to create billion+ graphs constrained by the physical 1-2 TByte RAM limits on a single node without having to distribute it . Distributed graph requires the queries and the solves to be performed by a more sophisticated algorithm that might need to jump back and forth between the sub-graphs (partitions) in an iterative manner until convergence. In the case of graph queries (adjacency traversals), the query algorithm might need to ping-pong among multiple partitions many times until the traversal reaches to the requested number of hops.

\begin{figure*}
\centering
    \framebox{\includegraphics[width=0.98\linewidth, keepaspectratio]{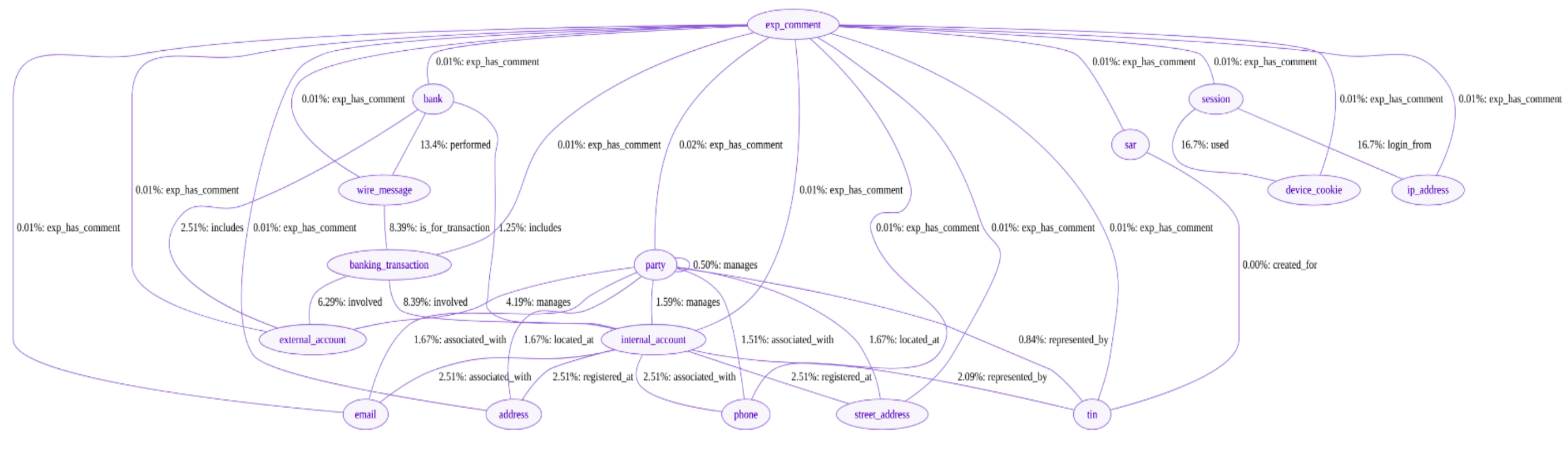}} 
    \caption{Automatic graph ontology generated from the node/edge labels - the percent values are indicating the proportion of number of edges between any two node labels over the entire graph. This graph has 4.3 billion edges and 2.8 billion nodes with 34 edge and 16 node labels.}
    \label{Figure:ontology}
\end{figure*}

\begin{figure}
\centering
    \includegraphics[width=0.9\linewidth, keepaspectratio]{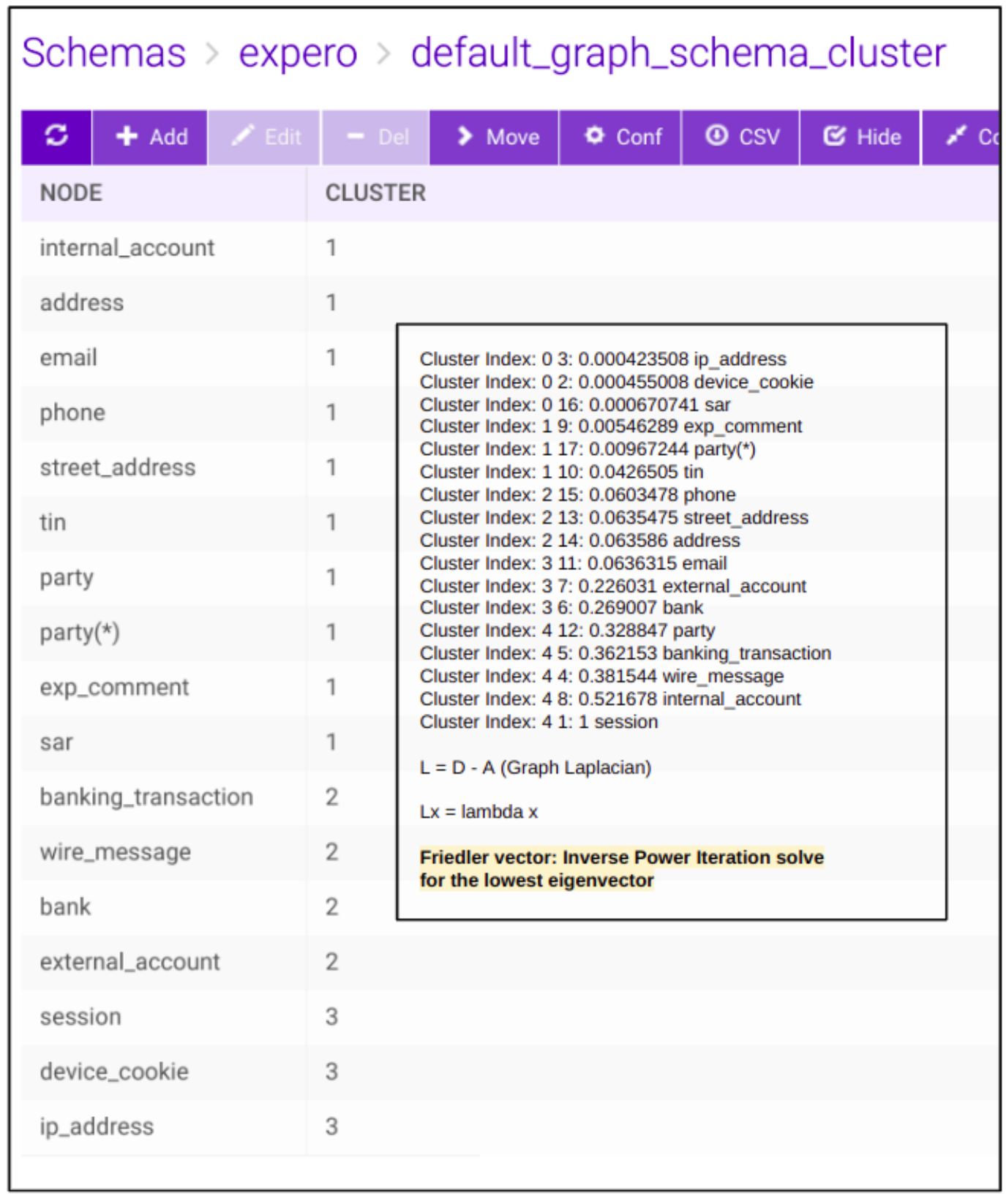} 
    \caption{Louvain (left) and Recursive Spectral Bisection (RSB) cluster partitioning (right) solvers applied over the label graph (ontology graph); each node label is associated with a cluster index. Both solvers are implemented in-house and freely available in Kinetica-Graph under match/graph endpoint.}
    \label{Figure:clusters}
\end{figure}

Therefore, when the graph size is too big to fit into the available physical memory, there is no other way but to distribute the graph which not only requires to have an effective partitioning scheme but also increases the total query/solve time substantially. Though graph partitioning is not the topic of this paper, and discussed in Kinetica-Graph paper~\cite{kineticagraph} in more detail, in short, our distribution topology does not require replication of edges but only the nodes at partition boundaries, known as duplicated nodes. The predicates for graph partitioning can be random, id based or geometrical (bounding boxes). However, with this labeling framework, we have come up with a new criterion for partitioning using the Louvain clustering algorithm~\cite{louvain} or recursive spectral bisection (RSB) algorithm~\cite{rsb} over the label graph (graph ontology) which is generated automatically while creating the main graph from the label connections. Louvain or RSB solvers are used to compute the clusters composed of labels so that the chosen cluster quality metric is maximized as seen in Figure~\ref{Figure:ontology} and Figure~\ref{Figure:clusters}. The respective SQL statement for creating this graph is also shown in Figure~\ref{Figure:creategraph}.  

\begin{figure}
\centering
    \framebox{\includegraphics[width=\linewidth, keepaspectratio]{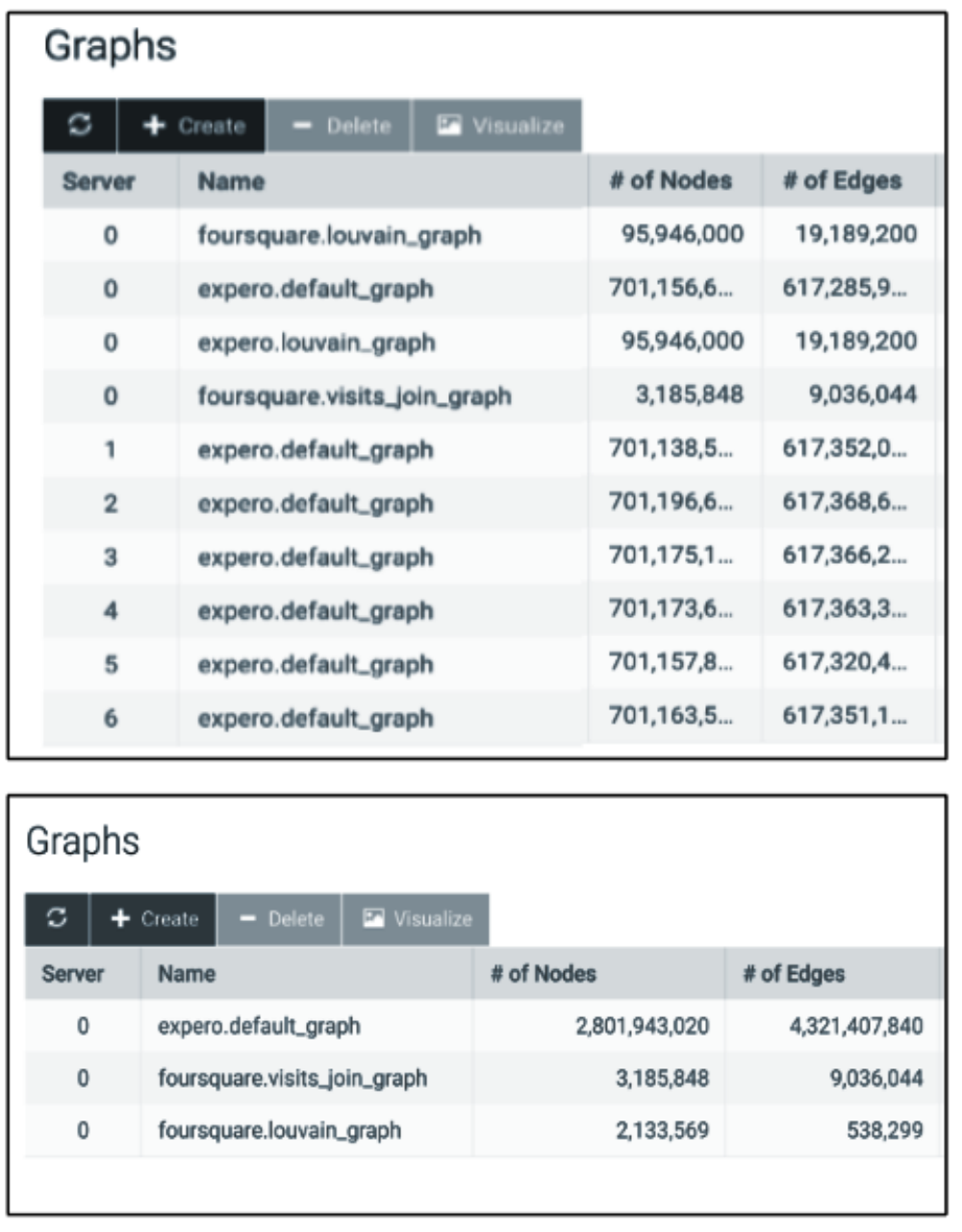}} 
    \caption{Distributed graphs 'expero.default\_graph' over six nodes of a hybrid Kinetica cluster (top), non-partitioned graph 'expero.default\_graph' on a single node (bottom) - A total of 4.3 billion edges and 2.8 billion nodes with 34 edge and 16 node labels. Each node and edge is associated with one or more labels.}
    \label{Figure:distributed}
\end{figure}

\section{Discussion and Conclusions}
\label{Section:Conclusions}

\begin{figure*}
\centering
     \framebox{\includegraphics[width=0.7\linewidth, keepaspectratio]{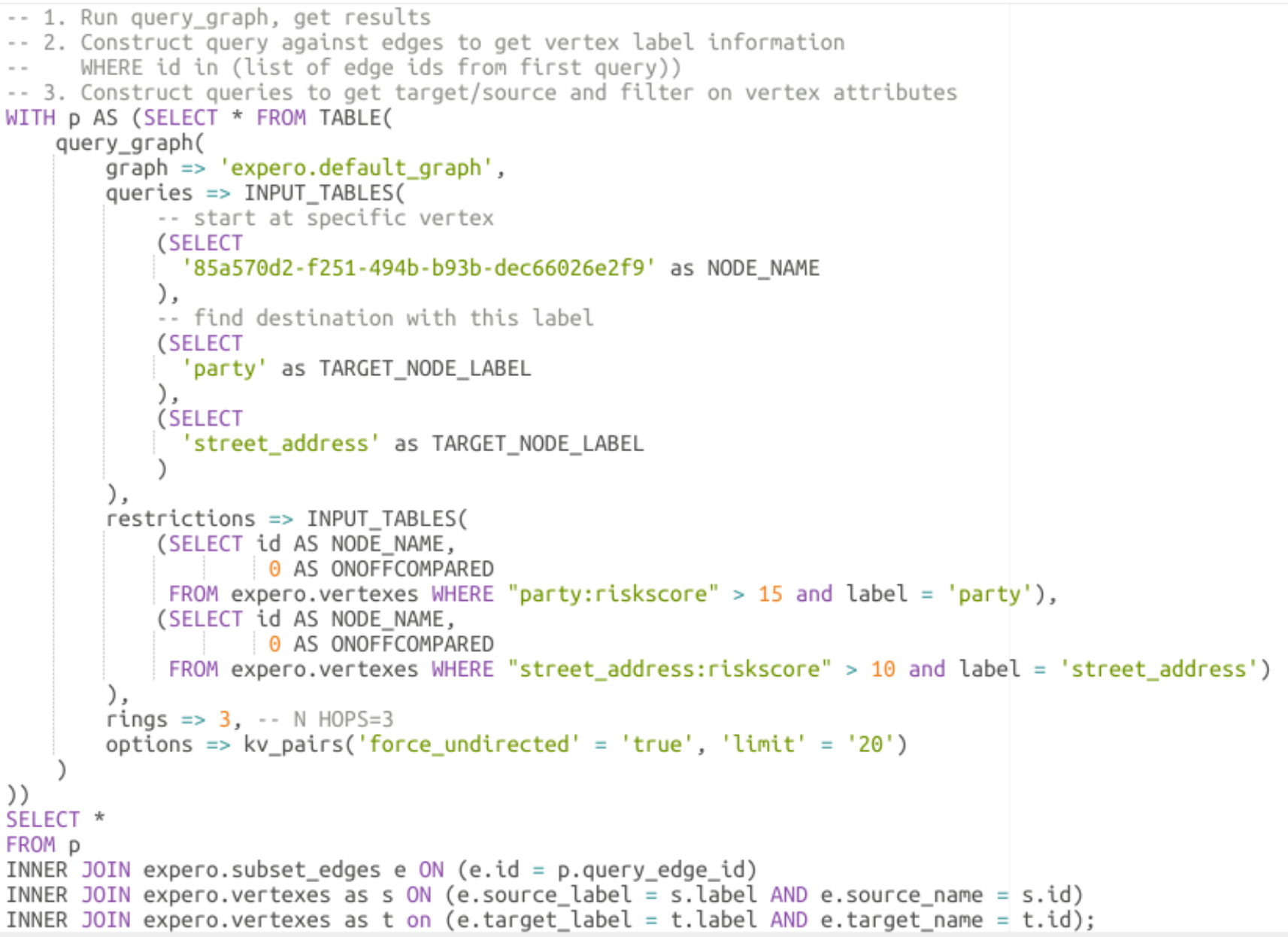}}
    \caption{Finds paths from source to target within N hops (n hop query) - source is always a known vertex, and the target is usually a set of match criteria (e.g. may not be a single vertex or a node label). N can range from 1 to however high we configure the UI to allow. Users usually want to go $3-5$ hops. In this example, the query aims to find $party$ or $street\_address$ vertexes whose $risk\_scores \le 15$ and $10$, respectively within 3 hops of the starting node. Note that the outer join operations are using the small graph query output of only 20 path long (max $\approx~100$ records) versus the 2.8 billion record original vertexes table.}
    \label{Figure:queries}
\end{figure*}

The best and the most efficient use case for the hybrid  Kinetica-Graph technology is the ability to solve the nearly impossible OLAP joins between large database tables. This problem is the most troublesome in the industry since the structured data and the  stenciled join operations require huge  memory foot prints due to the need for \textit{squared} dimension of these tables. Hence, the championing idea is having the unstructured graph solves in the inner section of the SQL queries to provide smaller result tables for the outer enclosing OLAP joins so that the joins could be done easily with much smaller result tables against the large database tables in order to pull in the other necessary attributes as the desired output. An example of this type of hybrid querying approach is shown in Figure~\ref{Figure:queries}. 

The 3-hop adjacency queries over distributed graphs for a total of 4.3 billion edges take about 5-8 seconds due to ping-ponging between the graph servers on the 6 partitions that are created using random-sharding on edge ids. The query speed is much improved using a non-partitioned single graph within 1.2-1.8 seconds using the conventional (old) key-value look-up labeling mechanism. However, using the proposed novel labeling data-structure, not only we were able to fit the giant graph within the physical memory limit of a 1TByte node but also get the query time under a second (400-700 miliseconds). If bigger graphs are needed, a portion (up-to $90\%$) of the physical RAM could have been used in ZRAM allocation with x4 compression with a very acceptable trade-off in speed within 20-40\% slow-down considering the other alternative as  distributed. Nevertheless, much slower distributed graph alternative is inescapable in order to stay completely scalable to any graph size, like reaching to peta scale size graphs, for example. However, for practical purposes, this new and novel labeling data-structure gives us the ability to be the lowest (possibly) on the memory requirement so that we could entertain fast queries within a single box up-to perhaps 20-40 billion graphs depending on the RAM (and ZRAM) limits.

\begin{figure}
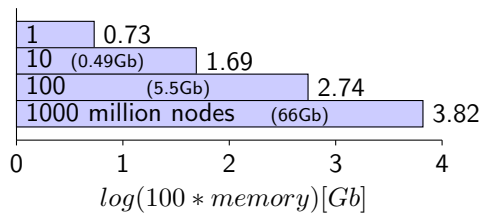

\centering
\begin{bchart}[step=1,max=4,scale=0.7]]
\bcbar[text=1]{0.73}
\bcbar[text=10~~\scriptsize{(0.49Gb)}]{1.69}
\bcbar[text=100~~~~~~~~~\scriptsize{(5.5Gb)}]{2.74}
\bcbar[text=1000 million nodes~~~~\scriptsize{(66Gb)}]{3.82}
\bcxlabel{ $log(100*memory) [Gb]$}
\end{bchart}
 \caption{Scalable linear memory consumption of the labeling infrastructure vs graph size. The memory impact of labeling datastructures for 1.2 billion node/edge graph is roughly 100 GBytes.}
    \label{Figure:memory}
\end{figure}

Another very important impact of this novel data-structure is such that the amount of storage is a constant fixed amount regardless of the variable number of labels that can be tethered to the nodes and edges - thanks to the in-place single linked lists embedded in one fixed size vector of twice the size of the number of nodes/edges. In other databases, having a set of super nodes/edges where high number of labels associated with particular nodes/edges can bring the operations to a halt or require very skewed allocations and large memory foot-prints due to expanding and shrinking number of associations. This presents no issues with our labeling framework particularly when there are label attachments for the majority of the graph nodes/edges. The required vector size is known and hence the memory can be reserved even before attaching the labels dynamically. Additional memory requirement for other data-structures such as the label to tuple index map and FIFO vector for recycling the tuple indexes explained in Section~\ref{Section:NewDataStructure}, is negligible because even in the billion+ graphs there are at most hundreds of distinct labels. The memory required for the entire labeling framework,  dominantly the in-place single linked list is linearly scalable against the graph size as shown in Figure~\ref{Figure:memory}. 

The future work on this novel labeling framework would be developing efficient partitioning algorithms possibly over the graph ontology (label graph) generated automatically during the ingestion process from database tables into generating the distributed graphs. We are also planning to adopt hybrid queries i.e., graph results as input for joins against large tables as our standard approach and provide viable and acceptable SLAs for the practical needs of our clients particularly in banking and retail industries. 
 
\section*{Acknowledgement}

The authors would like to thank Expero, who provided a descriptive graph data-set to simulate a real world credit card risk scoring use case so that we could improve Kinetica-Graph's labeling framework explained in this manuscript.

\section*{Notes on Contributors}
\small{
\noindent \textbf{Bilge Kaan Karamete} is the lead technologist for the Geospatial and Graph efforts at Kinetica. His research interests include computational geometry/algorithm development, unstructured mesh generation, parallel graph solvers. He holds PhD in Engineering Sciences from the Middle East Technical University, Ankara Turkey, and post doctorate in Computational Sciences from Rensselaer Polytechnic Institute, Troy New York.

\noindent \textbf{Eli Glaser} is VP of Engineering at Kinetica. He leads the development teams concentrating in data analytics, query capability and performance. Eli holds Master's in Electrical Engineering from The Johns Hopkins University, Baltimore Maryland.
}

\section{Software avaliability}

The labeling framework and data-structures written as header only c++ code discussed in this manuscript is publicly available in the github repository of \textbf{kineticadb/kinetica\_graph\_labels}. Kinetica and Kinetica-Graph is also freely available in Kinetica's Developer Edition at \textbf{https://www.kinetica.com/try} that the use cases depicted in this manuscript can easily be replicated by the readers.

\section*{References}

\bibliography{label}

\begin{thebibliography}{10}
\expandafter\ifx\csname url\endcsname\relax
  \def\url#1{\texttt{#1}}\fi
\expandafter\ifx\csname urlprefix\endcsname\relax\def\urlprefix{URL }\fi
\expandafter\ifx\csname href\endcsname\relax
  \def\href#1#2{#2} \def\path#1{#1}\fi

\bibitem{stl}
D.~R. Musser, A.~Saini, The STL Tutorial and Reference Guide: C++ Programming
  with the Standard Template Library, Addison Wesley Longman Publishing Co.
  USA, 1995.

\bibitem{cppdata}
M.~A. Weiss, Data Structures and Algorithm Analysis in C++, Addison Wesley
  Longman Publishing Co. Boston USA, 1998.

\bibitem{graphdbs}
D.~Fernandes, J.~Bernardino, Graph databases comparison: Allegrograph,
  arangodb, infinitegraph, neo4j, and orientdb., in: Data, 2018, pp. 373--380.

\bibitem{neo4j}
J.~Guia, V.~G. Soares, J.~Bernardino, Graph databases: Neo4j analysis., in:
  ICEIS (1), 2017, pp. 351--356.

\bibitem{tiger}
A.~Deutsch, Y.~Xu, M.~Wu, V.~Lee, Tigergraph: A native mpp graph database
  (2019).
\newblock \href {http://arxiv.org/abs/1901.08248} {\path{arXiv:1901.08248}}.

\bibitem{networkx}
A.~A. Hagberg, D.~A. Schult, P.~J. Swart, Exploring network structure,
  dynamics, and function using networkx, in: G.~Varoquaux, T.~Vaught,
  J.~Millman (Eds.), Proceedings of the 7th Python in Science Conference,
  Pasadena, CA USA, 2008, pp. 11 -- 15.

\bibitem{rbt}
R.~Sedgewick,
  \href{https://api.semanticscholar.org/CorpusID:199513342}{Left-leaning
  red-black trees}, in: Left--leaning Red-Black Trees, 2008, pp. 1--10.
\newline\urlprefix\url{https://api.semanticscholar.org/CorpusID:199513342}

\bibitem{rbt2}
S.~Sainz-Palacios, Flat combined red black trees (2019).
\newblock \href {http://arxiv.org/abs/1912.11417} {\path{arXiv:1912.11417}}.

\bibitem{hash}
T.~Ozsari, A hash of hash functions (2003).
\newblock \href {http://arxiv.org/abs/cs/0310033} {\path{arXiv:cs/0310033}}.

\bibitem{hash2}
D.~Köppl, Separate chaining meets compact hashing (2019).
\newblock \href {http://arxiv.org/abs/1905.00163} {\path{arXiv:1905.00163}}.

\bibitem{probing}
M.~Thorup, Linear probing with 5-independent hashing (2017).
\newblock \href {http://arxiv.org/abs/1509.04549} {\path{arXiv:1509.04549}}.

\bibitem{probing2}
D.~E. Knuth, The Art of Computer Programming, Volume 1 (3rd Ed.): Fundamental
  Algorithms, Addison Wesley Longman Publishing Co., Inc., USA, 1997.

\bibitem{benchmark}
M.~Leitner-Ankerl, Comprehensive c++ hashmap benchmarks 2022,
  \url{https://martin.ankerl.com/2022/08/27/hashmap-bench-01}, accessed:
  2022-08-27.

\bibitem{kineticagraph}
B.~K. Karamete, L.~Adhami, E.~Glaser, \href{https://arxiv.org/abs/2201.02136}{A
  fixed storage distributed graph database hybrid with at-scale olap expression
  and i/o support of a relational db: Kinetica-graph} (2022).
\newblock \href {http://dx.doi.org/10.48550/ARXIV.2201.02136}
  {\path{doi:10.48550/ARXIV.2201.02136}}.
\newline\urlprefix\url{https://arxiv.org/abs/2201.02136}

\bibitem{graphviz}
GraphViz, Graphviz, \url{https://graphviz.gitlab.io/}, accessed: 2023-10-02.

\bibitem{dls}
B.~K. Karamete, R.~Aubry, E.~L. Mestreau, S.~Dey, A novel double link structure
  (dls) with applications to computational engineering and design, AIAA
  Aerospace Sciences Meeting 54 (2016) 1301.
\newblock \href {http://dx.doi.org/10.2514/6.2016-1301}
  {\path{doi:10.2514/6.2016-1301}}.

\bibitem{zram}
G.~F. Oliveira, et~all, \href{https://arxiv.org/abs/2111.02325}{Extending
  memory capacity in consumer devices with emerging non-volatile memory: An
  experimental study}, CoRR abs/2111.02325.
\newblock \href {http://arxiv.org/abs/2111.02325} {\path{arXiv:2111.02325}}.
\newline\urlprefix\url{https://arxiv.org/abs/2111.02325}

\bibitem{louvain}
V.~D. Blondel, J.-L. Guillaume, R.~Lambiotte, E.~Lefebvre, Fast unfolding of
  communities in large networks, Journal of Statistical Mechanics: Theory and
  Experiment 2008~(10) (2008) P10008.
\newblock \href {http://dx.doi.org/10.1088/1742-5468/2008/10/p10008}
  {\path{doi:10.1088/1742-5468/2008/10/p10008}}.

\bibitem{rsb}
Y.~Hu, R.~J. Blake, Numerical experiences with partitioning of unstructured
  meshes, Parallel Computing 20 (1994) 815--829.

\end{thebibliography}

\end{document}